\documentclass[preprint,showpacs,preprintnumbers,amsmath,amssymb,prb]{revtex4} 
\usepackage{graphicx}   
\usepackage{epsfig}
\usepackage{latexsym}

\begin{document} 
\title{Coverage dependence of the 1-propanol adsorption on the Si(001) surface and fragmentation dynamics} 
\author{Jian-Ge Zhou, Frank Hagelberg}   
\affiliation{Computational Center for Molecular Structure and Interactions, 
Department of Physics, Atmospheric Sciences, and General Science, Jackson 
State University, Jackson, MS 39217, USA }  
\author{Chuanyun Xiao}   
\affiliation{Department of Chemistry, New York University, New York, NY 10003, USA }  
  
\begin{abstract}
The geometric, electronic, energetic, and dynamic properties of 1-propanol adsorbed on 
the Si(001)-$(2\times 1)$ surface are studied from first principles by use of a slab 
approach. The 1-propanol molecule initially interacts with the Si surface through formation of a 
dative bond, subsequently the physisorbed 1-propanol molecule reacts with the surface by cleavage of the 
O-H bond, and the Si(001)-$(2\times 1)$  surface undergoes further reconstruction as a result of the adsorption 
of the organic species. The band structure and density of states (DOS) are first analyzed for 
this system. The band gap of the Si/1-propanol film increases as the coverage level is enhanced.
Good agreement is found with available experimental data.

\end{abstract}  
\maketitle   
\section{Introduction} 

The chemisorption of organic molecules on silicon surfaces is a highly topical 
subject of current research, both experimental and computational.
This interest may be ascribed to both the fundamental nature of this problem, 
involving the interaction between finite units 
and periodic substrates, but also to its relevance to various areas of recent technology, such as insulator films, nanolithography, chemical and biological sensors, and molecular electronics. 
The organic layers are formed by depositing organic compounds on the semiconductor surface. 
In order to optimize this process, the understanding of the interaction between the surface and the organic 
species is crucial. The majority of the reactions between the semiconductor surface and organic molecules 
occur at or near the dangling bonds of the reconstructed surface. 
For a silicon (001) surface, the $2\times 1$ reconstruction  
leads to the formation of silicon dimers, where a strong $\sigma$  bond and a weak $\pi$ bond between the two dimer atoms is 
observed\cite{chd}.  It is well known that for hydrocarbons, the C-C double bonds break the dimer $\pi$ bond and lead to 
the formation of the new surface bonds that are energetically favorable\cite{hhgp}. Employing a similar mechanism, 
one can produce well ordered organic films/Si structures with a stable and uniform interface. These composites of silicon surfaces coated by organic films may lead to novel types of microelectronic 
devices that exploit the rich variety of functional groups of the organic species.

In the past decade, the reaction between the silicon surface dimers and alcohols have attracted 
much attention\cite{ebm}$^{-}$\cite{zcc}. For instance, 
the adsorption of ethanol on Si(001) was first observed by using surface infrared absorption 
spectroscopy\cite{ersc}. 
At room temperature, the ethanol is adsorbed dissociatively to form surface bound hydrogen and ethoxy groups, 
as a consequence of O-H bond breaking. 
The adsorption of ethanol on Si(001) at room temperature has 
also been studied employing high-resolution synchrotron radiation photoemission\cite{czcp}. In this case, O-H bond 
scission occurs. This behavior is at variance with ethanol adsorption on 
Si(111)-($7\times 7$)\cite{cppw}, where the C-O bonds where found to be broken.
The reaction of 1-propanol ($C_{3}H_{8}O$) with the Si(001)-($2\times 1$) surface was investigated in the pioneering work of Zhang et al. \cite{zcc} by Auger electron 
spectroscopy and thermal desorption spectroscopy. From this study, the 1-propanol molecule 
initially interacts with the Si surface through the 
formation of a ``dative bond", followed by further reaction of the physisorbed 1-propanol molecule 
with the surface by O-H bond cleavage. 
From the work reported in reference \cite{zcc}, the O-H bond 
cleavage is a kinetically favored reaction, but the O-C bond cleavage is thermodynamically preferred. 

So far, there is no full first-principles theoretical calculation that provides a complete
description of the 1-propanol molecule reaction with the Si(001) surface, excepting a preliminary calculation in 
the framework of a single-dimer cluster model for the Si(001) surface\cite{zcc}. Adopting the latter model, 
however, the surface-specific aspects of the problem at hand cannot be treated adequately. 
Thus, it is impossible to simulate the ``buckling" of the surface dimer. 
As mentioned in the work of Zhang et. al. \cite{zcc}, the results based 
on a single-dimer model should be substantially 
improved by considering an array of dimer clusters to account adequately for charge delocalization or 
surface relaxation phenomena. Secondly, the electronic properties, such as the band structure and the density 
of states (DOS) distribution 
for this system, have not been discussed before. For in-depth analysis of the substrate-adsorbate interaction, 
however, the understanding of these features is of crucial importance. Thirdly, studying the dependence
of various characteristic properties on the 1-propanol coverage is hardly feasible in the framework of a 
single-dimer model\cite{xhl}. Finally, the mechanism of the reaction between the 1-propanole molecule 
and the Si(001) surface 
has not been studied before. Such a simulation, involving the interaction of the finite 
molecular adsorbate and the periodic substrate at room temperature has been performed in the context of the 
present work by means of ab initio molecular dynamics (MD), as described in further detail below.

Guided by this motivation, in the present contribution we study the adsorption of 
1-propanol on the Si(001)-($2\times 1$) surface 
by use of the VASP code\cite{khf}, involving a slab geometry and periodic boundary 
conditions. The introduced model allows for an appropriate 
description of the Si(001) surface with and without the 
adsorption of the 1-propanol molecule as the reconstruction of the Si surface before and 
after adsorption can be displayed manifestly.
Section III of this contribution contains a detailed analysis of the most prominent reactions 
undergone by 1 - propanol on the Si(001) surface, including the calculations of the reaction barriers 
corresponding to various reaction pathways. Further, the charge density in a plane including
the 1 - propanol oxygen and a surface silicon atom, the
surface band structure within the silicon fundamental gap, and
the DOS and partial DOS distributions projected on the substrate atoms as well as the
1-propanol molecule or its fragments are discussed.
In addition, we outline the variation of the binding energies, the energy barriers, the DOS, 
and the energy gap  with the degree of coverage, where four coverage levels (0.125 ML, 0.25ML, 0.5ML, and 1.0ML)
are taken into account.
Finally, we will make admission for finite temperature and compare 
the characteristic reaction mechanisms at T = 300 K with those found at T = 0 K.

\section{Computational method}  

Our calculations were carried out by use of the VASP code\cite{khf}. Density functional theory (DFT) was applied 
on the level of the generalized gradient approximation (GGA)\cite{pw}  in conjunction with the PAW 
potential\cite{kj}$^{,}$ \cite{peb}. 
The wave functions are expanded in a plane wave basis with an energy cutoff of 400 ev, whereas the cutoff
for the augmentation charges is 645 ev. The Brillouin zone integrations 
are performed by use of the Monkhorst-Pack scheme\cite{mp} with the origin shifted to the $\Gamma$ point. 
We utilized a $3\times 3\times 1$  k point mesh for the geometry optimization, and $8\times 8\times 1$  k point
mesh for the DOS calculation.
The Si(001)-($2\times 1$)  surface is modeled adopting a supercell geometry with an atomic slab of 5 Si 
layers where terminating hydrogen atoms passivate the Si atoms. 
The supercell consists of a $4\times 4$ ideal cell, i.e. 80 atoms and 32 H atoms. The Si atoms in the 
top four atomic layers are allowed to relax, 
while the Si atoms in the bottom layer and the adjacent passivating H atoms are fixed to simulate 
bulk-like termination\cite{csb}$^{-}$\cite{scb}. 
The vacuum region is about 19 atomic layers, which 
exceeds the length of the 1-propanol molecule and provides sufficient 
spacing for the present MD simulation.
We performed computations on the pure substrate that were intended to examine 
the accuracy of our approach. Thus, we increased the energy cutoff to 500 eV and the number of 
k points to $8\times 8\times 1$. Neither of these tests led to any appreciable
changes of total surface energy; in both cases, the difference amounted
to less than $1.2\%$.
The energy barriers characterizing different reaction paths were calculated by the 
"climbing" Nudged Elastic Band (NEB) \cite{huj}$^{-}$\cite{jmj} method with six images, 
which permits identifying minimum energy paths 
in complex chemical reactions. Ab initio Molecular Dynamics simulations were performed by use of 
a Verlet algorithm to integrate Newton's equations of motion. The canonical ensemble was simulated
using the Nos\'{e} algorithm \cite{nose}. 

\begin{figure}
\includegraphics[width=4.4in]{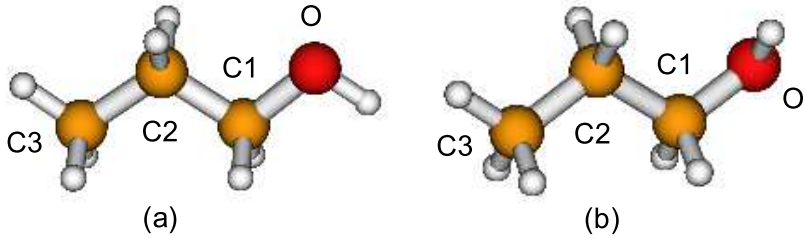}
\vspace{-0.5cm}
\caption{(Color online) 1-propanol molecule structures in the gas-phase.}  
\label{propan}
\end{figure}

As a test, we calculated the structural properties of the free 1-propanol molecule, and found the obtained bond 
lengths to be in good agreement with the respective findings\cite{dd}$^,$\cite{nist}. 
The deviation from these earlier results 
was found to be less than three per cent. The structures of the isolated 1-propanol molecule are 
shown in Fig.\ref{propan}. Further, the calculated 1-propanol ionization energy is within eight per 
cent of the experimental value, 10.18 $\pm$ 0.06 eV\cite{hb}

The calculated geometrical parameters are given in Table \ref{structure}, where the unit of the bond 
length is $\AA$.

\begin{table}
\caption{The calculated structural parameters of the isolated two isomers of the 1-propanol molecule}
\begin{center}
\begin{tabular}{ccc}
\hline
~ &  Structure (a)~~ & Structure (b)\\
\hline
$d(O-H)$~~~      & 0.97    & 0.97\\
$d(O-C1)$      ~ & 1.44    & 1.43\\
$d(C1-C2)$     ~ & 1.52    & 1.52\\
$d(C2-C3)$  ~~   & 1.53    & 1.53\\
$\angle H-O-C1$  & 108.5   & 108.0\\
$\angle O-C1-C2$~ & 108.3 & 113.5 \\
$\angle C1-C2-C3$~ & 112.7 & 112.7 \\
$Dihe(H-O-C1-C2)$~ & 179.9 & 61.3\\
\hline
\end{tabular}
\end{center}
\label{structure}
\end{table}

It should be noted that there are actually five conformers for the 1-propanol molecule which differ from each 
other with respect to the  dihedral angles. The isomers (a) and (b) as shown in Fig.\ref{propan} are the 
energetically favored species\cite{tsy}, 
and these two conformers are readily interchanged at room temperature, since the OH torsion barriers are quite low. 
Our calculations involve the structure (a) which deviates from structure (b) by a difference in binding 
energy lower than 0.01 ev.

\begin{figure} 
\begin{center}
\includegraphics*[width=4.5in]{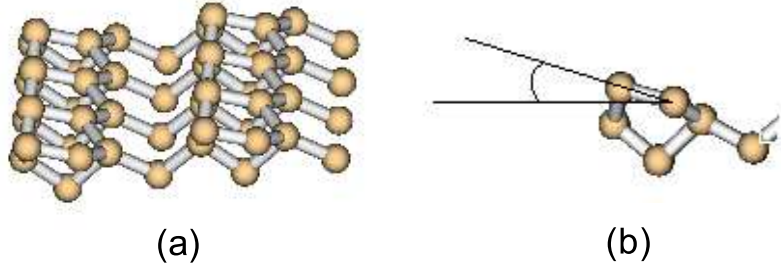}
\vspace{-0.5cm}
\end{center}
\caption{(Color online) The $2\times 1$ reconstructed silicon surface. The three top layers are shown.}  
\label{1x2}
\end{figure}

We further computed energetic and geometric parameters pertaining to the $2\times 1$ 
reconstruction of the bare Si(001) surface (its explicit illustration
can be found from Fig. 1 in reference \cite{rbk}). 
The $2\times 1$ reconstructed silicon surface is displayed in Figure \ref{1x2}. 
Figure \ref{1x2}a reveals the three silicon top layers, and
Figure \ref{1x2}b illustrates the buckling angle, which is defined as the angle 
between the dimer row  and the horizontal plane. The Si dimers are oriented along the x axis or $[110]$
direction, and the dimer rows are along the y axis or the $[\bar{1}10]$ direction. 

For the $2\times 1$ surface reconstruction with asymmetric Si dimers, the energy gain is 1.6 ev per dimer. 
The internuclear distance between the two Si centers is 2.32 $\AA$. The distance between two 
adjacent dimers perpendicular
 to the row is 3.86 $\AA$. The distance between the ``up" Si atom of one dimer to the ``down" Si atom of the 
next is 5.57 $\AA$. The buckling angle is 18.0$^{\circ}$. These results agree with existing experimental\cite{owr} 
data and other calculations\cite{csb}.  

\section{Results and discussion}

\subsection{The physisorbed and chemisorbed configurations}  

In this section, we first describe the stable physisorbed configurations of the 1-propanol molecule 
on the Si(001)-($2\times 1$) surface for 0.125 ML, where three non-dissociative structures are identified. Subsequently, we 
consider seven dissociated structures which correspond to chemisorbed configurations. 
From these seven cases of chemisorption, we select the two most stable ones. Further, we characterize 
the modification of the bare Si(001)-($2\times 1$) reconstructed surface due to the physisorbed and chemisorbed 
1-propanol molecules. Moreover, in an effort to examine which one among the chemisorbed structures is 
most likely to be observed experimentally, we calculate the energy barriers relevant to the chemisorbed 
configurations. The charge density in the plane passing through the oxygen and silicon atom
is drawn to show how the O-Si bond is formed. 
The surface band structures,
the electronic density of states (DOS) and partial DOS projected on the Si atoms and the 1-propanol
molecule (or its fragments) are obtained to characterize the interaction between the substrate and the adsorbate.

\begin{figure} 
\includegraphics[width=4.0in]{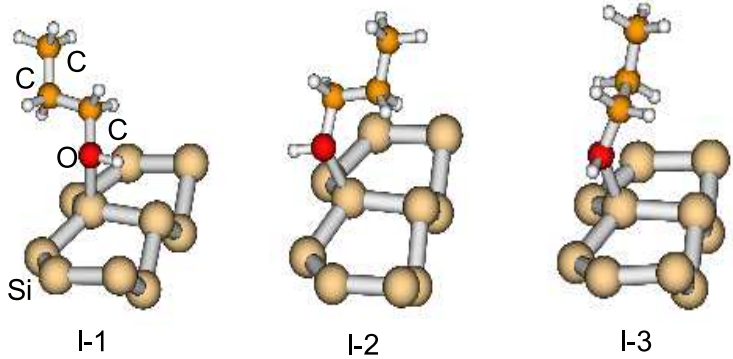}
\vspace{-0.5cm}
\caption{(Color online) Three stable physisorbed configurations of the 1-propanol molecule on the Si(001) surface.
The yellow (largest), red, orange, and 
white (smallest) spheres (from bottom to top) represent the silicon (Si), oxygen (O), carbon (C) and
hydrogen (H) atoms, respectively. To illustrate the interaction between 
the 1-propanol molecule and the Si(001)-($2\times 1$)  surface, we display
only a few surface Si atoms. Our simulation includes 16 Si atoms in each layer.}
\label{physisorption}  
\end{figure} 
 
Here we only focus on the adsorbed structures obtained by an exothermic process, i.e., the composite of 
the surface and the adsorbed species is lower in energy than the free 1-propanol molecule and the bare 
Si(001)-($2\times 1$) surface in separation from each other. Experimentally, it has been demonstrated that the 1-propanol 
molecule and its fragments are oriented vertically with respect to the surface\cite{zcc}. 
This adsorption geometry is 
therefore adopted for our treatment of the physisorbed and chemisorbed configurations.

Our calculated results confirm that the 1-propanol molecule initially interacts with the Si(001)-($2\times 1$)  
surface via the 
formation of a ``dative bond" between the oxygen atom and the electrophilic ``down" Si atom of the surface dimer. 
Specifically, the O - Si bond may be characterized as a covalent connection arising from the lone pair of the O atom. 
The 1-propanol molecule remains essentially intact (this motivates our nomenclature I-1, I-2 and I-3 for the 
physisorbed configurations) or undissociated on the physisorbed sites, and assumes various orientations 
of the O-H bond with respect to the Si surface. The obtained structures are shown in Figure \ref{physisorption} 
which illustrates that the direction of 
the O-H bond can be parallel (I-1), antiparallel (I-2) or perpendicular (I-3) to the Si dimer. However, the energies 
of the three configurations are very close to each other, i.e., the rotation of the 1-propanol 
molecule around the Si-O bond is 
quite facile. 

In the single-dimer cluster calculation, only the physisorbed structure similar to I-2 was considered \cite{zcc},
while we include three possible physisorbed configurations I-1, I-2 and I-3 here.
Table \ref{geometry} shows that the reported binding energy (0.39ev)\cite{zcc} is 
considerably smaller than that found in this work (0.72ev). 
This discrepancy might be attributed to the difference between the single dimer model and the periodic 
approach followed in the present approach. 

\begin{table}
\caption{The binding energies in eV/per 1-propanol molecule
and structural parameters of the configurations shown in Figs.\ref{physisorption} and 
\ref{chemisorption}. The data are for 0.125 ML 
coverage. In parenthesis, the binding energy values obtained by a single-dimer model are indicated.}
\begin{center}
\begin{tabular}{cccccccccccc}
\hline
~ & I-1&I-2&I-3&F-1&F-2&F-3&F-4&F-5&F-6&F-7&F-8\\
\hline
$E_{bind}$~&0.75~&0.72~&0.68~&2.59~&2.89~&1.54~&1.66~&1.66~&1.65~&2.22~&2.46~\\
~~&~~&(0.39)~&~~&(2.62)~&(3.22)~&(1.66)~&(1.71)~&(1.71)~&~&~\\
$d(O-H)$~&1.01~&0.98~&0.98~&-~&0.97~&0.98~&0.99~&0.97~&-~&-~&-~\\
$d(C1-O)$~&1.48~&1.49~&1.49~&1.44~&-~&1.44~&1.43~&1.43~&~1.28&-~&-~\\
$d(C1-C2)$~&1.51~&1.51~&1.51~&1.52~&1.53~&1.53~&1.52~&1.52~&1.49~&1.53~&1.53~\\
$d(C2-C3)$~&1.53~&1.53~&1.53~&1.53~&1.54~&1.53~&1.53~&1.53~&1.53~&1.54~&1.53~\\
$d(O-Si)$~&1.96~&1.95~&1.99~&1.66~&1.68~&-~&-~&-~&1.79~&1.57~&1.57~\\
$d(C-Si)$~&-~&-~&-~&-~&1.93~&1.95~&1.94~&1.91~&-~&1.91~&1.91~\\
$d(H-Si)$~&-~&-~&-~&1.50~&-~&1.50~&1.50~&1.50~&1.50~&1.50~&1.50~\\
$d(Si-Si)$~&2.39~&2.38~&2.40~&2.43~&2.44~&2.41~&2.41~&2.43~&2.41~&2.46~&2.44~\\
$Buckling (degree)$~&8.7~&11.0~&10.8~&1.9~&4.1~&4.5~&5.7~&1.6~&2.7~&1.3~&1.9~\\
\hline
\end{tabular}
\end{center}
\label{geometry}
\end{table}

For the bare Si(001)-($2\times 1$) surface,  the "buckling" angle with the horizontal plane is 18.0$^{\circ}$.  
As a consequence 
of 1-propanol physisorption, the buckling angles for I-1, I-2 and I-3  become 8.7$^{\circ}$, 11.0$^{\circ}$ 
and 10.8$^{\circ}$, respectively.  
For the adjacent Si dimer, the corresponding buckling angles are 17.9$^{\circ}$,  17.8$^{\circ}$, and 17.0$^{\circ}$.
As the latter values are close to the angle found for the bare Si(001) surface, 18.0$^{\circ}$, the interaction between the 1-propanol molecule and the adjacent Si dimer is quite weak.

\begin{figure} 
\includegraphics[width=4.5in]{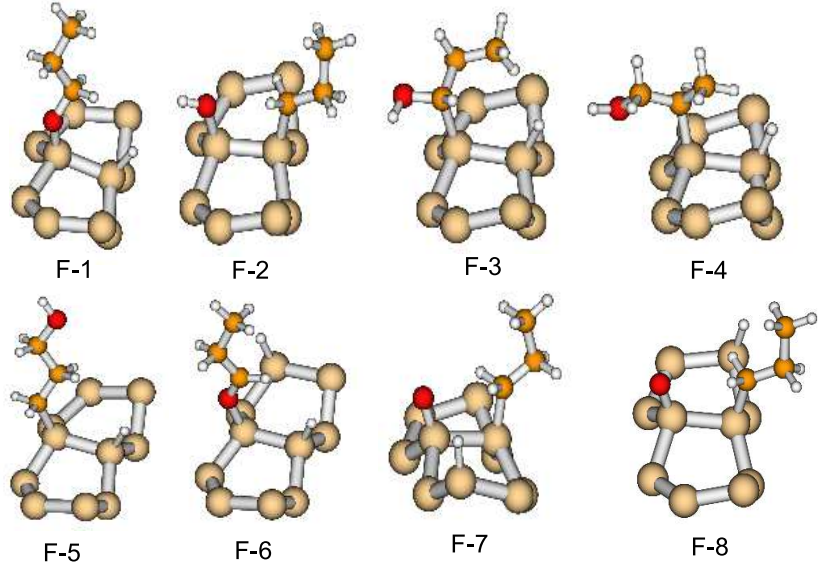}
\vspace{-0.5cm}
\caption{(Color online) The fragmented chemisorption structures of 1-propanol on the Si (001) surface.
For the sake of clarity, we have included only ten Si atoms in this illustration.}  
\label{chemisorption}
\end{figure}

We have verified that the physisorption in the cases I-1, I-2 or I-3 is a barrierless reaction, which starts from a 
1-propanol molecule far from the surface.  Once this is physisorbed and attached to the surface by a ``dative bond", 
the 1-propanol can proceed to react with the surface via a number of pathways, which break one or more molecular 
bonds to form dissociated configurations of increased stability. The eight 
principal dissociated structures arising from H atom loss or O - C bond cleavage are 
shown in Figure \ref{chemisorption}. 

The F-1 structure is obtained by breaking the O-H bond and detaching the H atom until it attaches to the ``up" Si 
atom of the same dimer to form a new H-Si bond.  This configuration has the second largest binding energy (2.59ev). 
The remaining alkoxy fragment is bonded to a Si surface dimer atom, while the separated H atom forms a bond with the other 
Si atom of the same dimer. The binding energy decreases as the H atom is attached to a Si atom of an adjacent dimer.  
For the ethanol molecule, the axis of the methyl group is almost perpendicular
 to the Si(001) surface\cite{ersc}, 
but in the case of 
1-propanol, the corresponding axis includes an angle of about $75^{\circ}$ degrees with the surface. Since the considered coverage of
 0.125 ML is sparse, the repulsion between the adjacent 1-propanol molecules is negligible. Thus we conclude 
that the vertical orientation of the 1-propanol molecule is not the result of the repulsion between alkoxy groups,  
which is consistent with the cases of ethanol adsorption\cite{ersc}. We have seen that the O atom forms a single polar 
covalent bond with only one Si atom of the surface dimer, which reflects the localized and directed nature of 
the Si dangling bond. 

The F-2 configuration is described by C1-O bond cleavage. The OH group and the alkyl fragment are bonded to the 
same Si dimer. If these two fragments are attached to adjacent dimers, the binding energy decreases. 
As Table \ref{geometry} shows, the F-2 configuration is thermodynamically most 
stable, i.e., it has the largest binding energy (2.89ev).  

The F-3, F-4 and F-5 configurations are characterized by breaking the C1-H, C2-H 
C3-H bonds, respectively, where the C1, (or C2, C3) atom is bonded to a Si atom and the detached H atom
forms a new bond with the other Si atom of the same dimer.  
From the respective binding energy one finds that the configuration of the C-H cleavage is
of lesser stability than both F-1 and F-2. 

To examine whether the F-1 and F-2 structures undergo further bond rupture, we consider
the configurations F-6, F-7 and F-8.
F-6 is described by the cleavage of a C-H bond in F-1, and the 
H atom is attached to the adjacent dimer. The energy of F-6 is higher than that of F-1 by 0.9 ev. 
F-7 and F-8 are obtained from F-2 by further dissociating the O-H bond and 
attaching the corresponding H atom to the 'down' and 'up' Si atom of the adjacent dimer.  
These structures are energetically less favored 
than the original F-2 configuration. 

In going from I-1 to F-1 (the H atom binding with the 'up' Si atom),  a substantial increase in
the binding energy is observed. The comparable transition from F-2 to F-7, however, is associated with a large decrease in the binding energy. 
This difference is related to the fact that I-1 is a physisorbed structure, while F-1 is a chemisorbed one, making plausible its higher stability as compared with I-1. Chemisorption is realized for the configuration F-2, involving saturated covalent bonding of the oxygen atom which forms one bond with the
'down' silicon atom and another one with the hydrogen atom. For the F-7 configuration, in contrast,  
the hydrogen atom is detached from the oxygen atom which consequently is unsaturated, implying a decreased 
binding energy for the F-7 structure. It may be assumed that oxygen in this configuration forms a double bond with the 'down' silicon atom. The latter, however, already forms two bonds with next layer silicon atoms, and a third one with its silicon dimer partner atom, which leaves a single bond between oxygen and silicon as the only possibility.
From an energetic point of view, F-1 and F-2 are most stable, 
corresponding to the tendency of 1-propanol to break the C-O bond or the O-H bond. 
Therefore, the subsequent discussion 
will be limited to the configurations F-1 and F-2.

To see that increasing the plane wave basis (or the energy cutoff) has only a slight effect on the above adsorption energies,
we have calculated the adsorption energies for the I-1, I-2, F-1 and F-2 configurations at 500eV.  
If the energy cutoff is 400ev, the adsorption energies for the same configurations are 0.75, 0.72, 2.59 and 2.89 eV, respectively. 
When the energy cutoff is 500ev, the corresponding energies are  0.75,  0.71,  2.56, and  2.91 eV.  
The difference between the two sets of results amounts to less that   $1.2\%$.    

\begin{figure} 
\begin{center}
\includegraphics[width=5.0in]{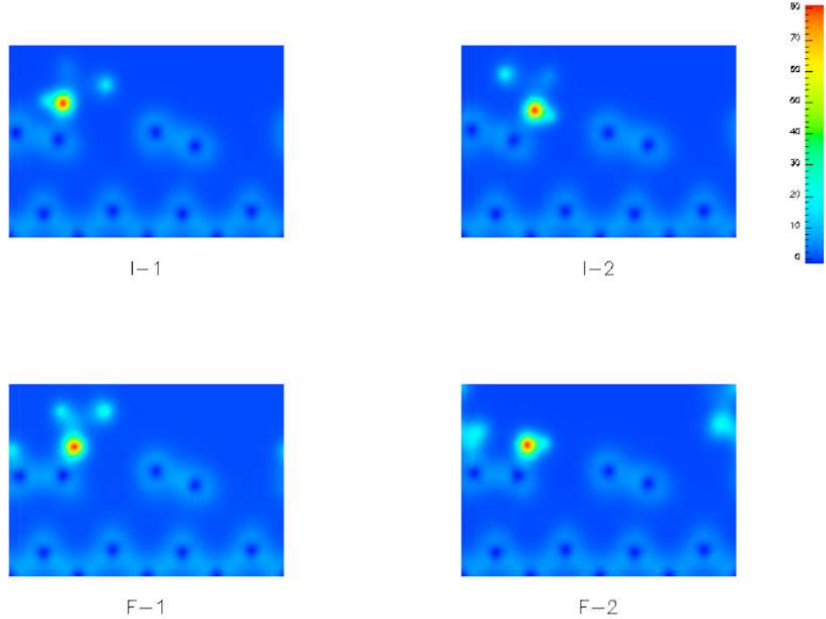}
\vspace{-1.1cm}
\end{center}
\caption{(Color online) The charge density in the plane extended in the directions [110] and [$\bar{1}$10]
for the configurations I-1, I-2, F-1 and F-2. The unit for the charge density is 
0.01 e/{$\AA^{3}$}. The red zone is due to the higher charge density
of the oxygen atom, the blue spots correspond to the silicon charge density. The charge
density between the O and Si atoms shows the polar O - Si bond.}  
\label{charge}
\end{figure}

The charge density in the plane passing through the oxygen and the silicon atom
(extended in the directions [110] and [$\bar{1}$10]) is shown in Fig.\ref{charge}.
It is calculated within pseudopotential framework.
The analysis of the charge density clearly shows that a polar ``dative bond" for the physisorbed structures 
I-1 and I-2 (O-Si bond lengths
for them are 1.96 $\AA$ and 1.95 $\AA$) and the polar covalent bond for chemisorbed structures
F-1 and F-2 (O-Si bond lengths
for them are 1.66 $\AA$ and 1.68 $\AA$) have been formed.  

\subsection{Energy barriers}  

To assess which chemisorbed structure is most likely to be observed experimentally, 
we have calculated the energy barriers relevant to the chemisorbed configurations. 
Table \ref{barrier} shows the energy barriers for the respective reactions.
For the physisorbed structures I-1, I-2 and I-3,  
the reaction proceeds without barrier.  For the cases of chemisorption, we have calculated the energy barriers 
for the processes that lead from I-1 to F-1, I-2 to F-2 and F-1 to F-6, respectively. The 
energy barriers have been calculated by the 
climbing "Nudged Elastic Band" method"\cite{huj}$^{-}$\cite{jmj}, where six 
equidistant images have been used. 

\begin{table}
\caption{The energy barriers $E_b$ , and transition state energy levels $E_{TS}$  with respect to the energy of 
the 1-propanol molecule and the Si(001).surface in separation from each other}
\begin{center}
\begin{tabular}{ccc}
\hline
Reaction~~~~~ &$E_b$(eV) ~~~~~ &$E_{TS}$(eV)~~~~~\\
\hline
1-propanol + Si(001) $\rightarrow$ I-1~~~~& 0~~~~~ & -~~~~~\\
I-1 $\rightarrow$ F-1 (O-H breaking)~~~~ & 0.05~~~~~ & -0.70~~~~~\\
I-2 $\rightarrow$ F-2 (C-O breaking)~~~~ & 1.34~~~~~ & 0.62~~~~~\\
F-1 $\rightarrow$ F-6~~~~~ & 2.9~~~~~ & 0.31~~~~~\\
\hline
\end{tabular}
\end{center}
\label{barrier}
\end{table}

For the transformation to the F-1 configuration, I-1 is the most favorable initial structure since its O-H bond is 
already oriented parallel to the Si dimer row. This reaction is a proton transfer process from oxygen to the 
electron-rich, nucleophilic "up" silicon atom of the dimer.  The I-1 to F-1 reaction is characterized by an 
energy barrier of 0.05 ev.  The binding energy of I-1 is 0.75 ev 
which implies that the barrier 
for the whole process, i.e. adsorption into the I-1 structure followed by transition to the F-1 structure, 
is below the initial energy, namely the of the free 1-propanol molecule and a bare Si(001) surface. Since the binding 
energy of F-1 is 2.59 ev, the I-1 $\rightarrow$ F-1 process is exothermic. 

\begin{figure} 
\begin{center}
\includegraphics[width=3.0in]{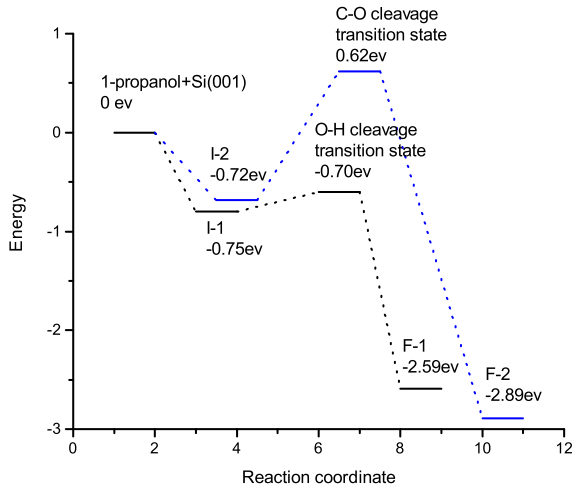}
\vspace{-0.9cm}
\end{center}
\caption{(Color online) The relative energy levels along the O-C and O-H cleavage reaction paths.}  
\label{transition}
\end{figure}

From the physisorption case I-2 to 
configuration F-2, involving the breaking of a C-O bond, the energy barrier is 1.34 ev. The binding 
energy of I-2 is 0.72. Therefore, the transition state energy is higher than the reference energy of 
the free 1-propanol molecule 
and the bare Si(001) surface. The binding energy of F-2 is 2.89 ev, making the I-2 $\rightarrow$ F-2 process 
exothermic too.  
Thus the O-C bond cleavage is thermodynamically stable, but the O-H bond cleavage is 
kinetically favored. In other words, the O-C bond cleavage has the highest binding energy, while the O-H bond 
cleavage has a smaller energy barrier than the O-C bond cleavage. This confirms, on the basis of a more adequate periodic model, the conclusion reached by Zhang et. al.\cite{zcc} in the framework of a finite cluster approach. The relative energies 
along the O-C and O-H cleavage reaction paths are schematically illustrated in 
Fig.\ref{transition}.

Zhang et. al.\cite{zcc} suggested that the initial O-H bond cleavage might be followed by a hydrogen elimination reaction 
to result in aldehydes and hydrogen.  Table \ref{barrier} shows that the energy barrier for the transition from the O-H 
cleavage configuration F-1 to the configuration F-6 is relatively high. One concludes that the respective 
reaction is not preferred. Similarly, the transition from F-2 to F-7 configuration is not favored. 

The undissociated structures I-1, I-2 and I-3 can be interpreted as metastable precursors for the more 
stable F-1 configuration. These precursors do not have sufficient binding energy at room 
temperature to compete as observable reaction products, i.e. the
cleavage of H is too fast for any of the physisorbed structures to be observed. The MD simulation outlined below gives additional 
support to this interpretation.

\subsection{Band Structure}  

\begin{figure} 
\begin{center}
\includegraphics[width=1.5in]{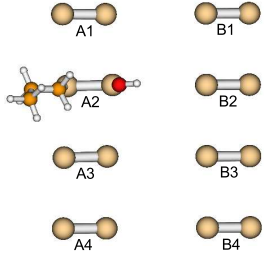}
\vspace{-0.9cm}
\end{center}
\caption{(Color online) The labels and positions of eight dimmer units in the top Si layer, where the horizontal 
direction is along [110] and
the vertical one is along [$\bar{1}$10].}  
\label{dimer}
\end{figure}

A sketch of the eight relevant Si dimer units in the top Si layer is shown in Fig. \ref{dimer} for unambiguous reference,
where the horizontal (vertical) corresponds to the [110] ([$\bar{1}$10]) directions, respectively. For
0.125 ML, the 1-propanol molecule or its fragments are adsorbed to the A2 dimer. 

The up Si atoms are located at the left dimer ends, 
the down Si atoms at the right. However, as described above, after the adsorption
of the 1-propanol molecule to the down Si atom, the latter is raised, i.e., the buckling angle decreases. 

\begin{figure} 
\begin{center}
\includegraphics[width=5.5in]{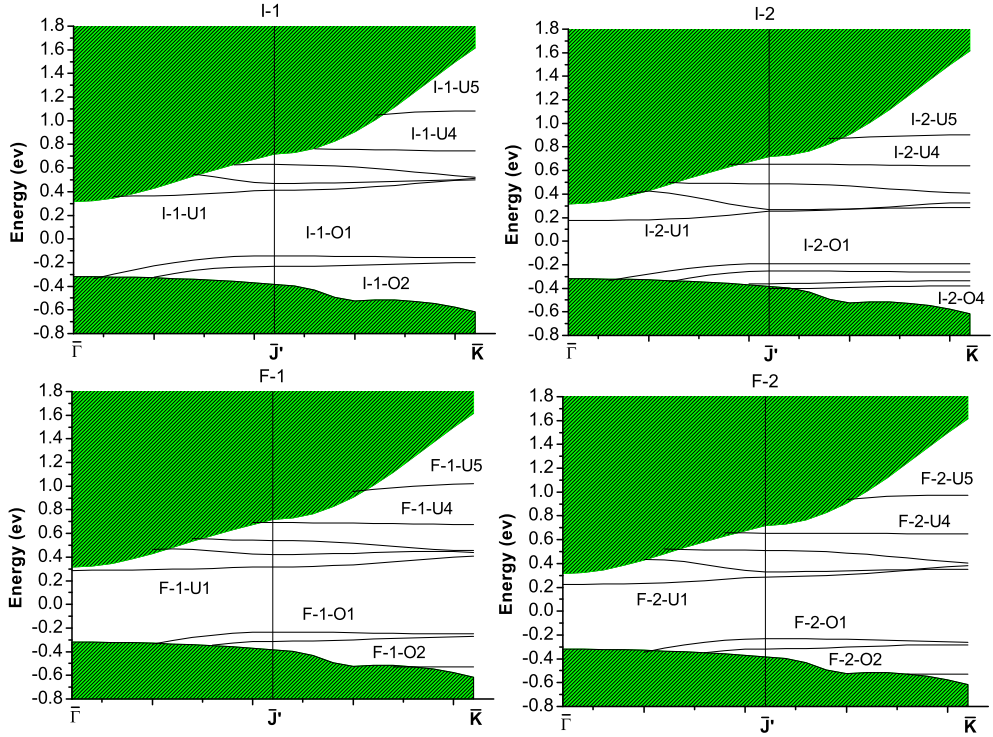}
\vspace{-1.5cm}
\end{center}
\caption{(Color online) Surface band structures for the configurations I-1, I-2, F-1 and F-2 at 0.125 ML.
The shaded areas represent the projected bulk band structure, while surface states are represented as solid lines.}
\label{band} 
\end{figure}

The surface band structures within the fundamental band gap of the silicon 
for the configurations I-1, I-2, F-1 and F-2 for 0.125 ML, are depicted in Fig. \ref{band}. The k points  
$\Gamma$,$J$,$K$,$J'$  are four vertices of the square of the quarter part
of the surface Brillouin zone (the relative positions of $\Gamma$,$J$,$K$,$J'$ points
can be seen from Fig.3 in the work of Ramstad et. al.\cite{rbk}).

Fig. \ref{band} reveals that there are seven, nine, eight and eight surface bands within 
the fundamental band gap of silicon for the
I-1, I-2, F-1 and F-2 configuration, respectively. The remaining valence (conduction)
bands lie in the lower (higher) shaded area. In the I-1 configuration,
for the valence bands (occupied), the top one is labeled I-1-O1, and the next lower one is I-1-O2, etc..
For the conduction band (unoccupied), the bottom one is referred to as I-1-U1, the higher ones are I-1-U2, I-1-U3, I-1-U4, and the
highest one is I-1-U5. The same nomenclature is used for the surface bands of the other three configurations. 

The two highest surface valence (O1 and O2) and the two lowest surface conduction bands (U1 and U2) contain the information
about the adsorption and are thus sensitive to the structural features of the surface. The conduction bands U4 and
U5, for instance, exhibit the same atomic orbital composition for all four configurations, I-1, I-2, F-1 and F-2. 
It is therefore sufficient to consider only the top two occupied valence bands (O1,O2) and bottom two
unoccupied conduction bands (U1,U2), and the
main contributions to these four bands are shown in Table \ref{bandcomp} for the configurations I-1, I-2, F-1 and F-2.
Table \ref{bandcomp} indicates that 
the band I-1-O1 contains the information about the 1-propanol adsorption. 
The A1 and A3 contributions to the valence band I-1-O1 are an electronic fingerprint of the adjacent dimers,
while the 1-propanol physisorption leaves their geometric structure unaffected, as reflected by their buckling
angles. 

\begin{table}
\caption{Atomic composition of the highest lying valence and lowest conduction bands for the configurations 
I-1, I-2, F-1 and F-2.
In parenthesis, the contributing atoms are indicated. }
\begin{center}
\begin{tabular}{cc}
\hline
Bands~~~~~~~ & main compositions\\
\hline
I-1-O1~~~~~~~ & $3p_z$ [$\it{up}$ Si (A1,A2,A3)]\\
I-1-O2~~~~~~~ & $3p_x$,$3p_x$ [$\it{up}$  Si (A1-A4,B1-B4)], $3p_z$ [$\it{up}$  Si (A4)]\\
I-1-U1~~~~~~~ & $3p_z$ [$\it{down}$ Si (A1,A3,A4,B1-B4)]\\
I-1-U2~~~~~~~ & $3p_z$ [$\it{down}$ Si (A1,A3,A4,B1-B4)] + $3p_z$ [Si in adjacent\\
~~~~~~~~~~~~~ & layer which form bonds with $\it{down}$ Si (A1-A4,B1-B4)]\\
\hline
I-2-O1~~~~~~~ & $3p_z$ [$\it{up}$  Si (A2)]\\
I-2-O2~~~~~~~ & $3p_z$ [$\it{up}$  Si (A1,A3)]\\
I-2-U1~~~~~~~ & $3p_z$ [$\it{down}$ Si (B1,B2,B3,B4)]\\
I-2-U2~~~~~~~ & 3s,3$3p_z$ [$\it{down}$ and $\it{up}$ Si (B1,B2,B3,B4)]\\
\hline
F-1-O1~~~~~~~ & $3p_z$ [$\it{up}$ Si (A1,A3,A4,B1,B3,B4)]\\
F-1-O2~~~~~~~ & 3s,$3p_z$ [$\it{up}$ Si (B2)]\\
F-1-U1~~~~~~~ & $3p_z$ [$\it{down}$ Si (A1,A3,A4,B1,B2,B3,B4)]\\
F-1-U2~~~~~~~ & $3p_z$ [$\it{down}$ Si (A1,A3)], 3s,$3p_z$ [$\it{up}$  Si \\
~~~~~~~~~~~~~ & (B2,B4)], $3p_z$ [$\it{down}$ Si (B2,B4)]\\
\hline
F-2-O1~~~~~~~ & $3p_z$ [$\it{up}$ Si (A4,B4)]\\
F-2-O2~~~~~~~ & 3s,$3p_z$ [$\it{up}$  Si (A1,A3,B1,B3)]\\
F-2-U1~~~~~~~ & $3p_z$ [$\it{down}$ Si (B1,B2,B3,B4)]\\
F-2-U2~~~~~~~ & $3p_z$ [$\it{down}$ Si (A1,A3)]\\
\hline
\end{tabular}
\end{center}
\label{bandcomp}
\end{table}

Here we note that within the fundamental band gap of silicon, there is no conduction band for
a Si-O bonding due to the adsorbate. In case of the acetonitrile adsorption on the silicon
surface\cite{mop}, in contrast, a conduction band with both Si and N  contributions is found within this gap, 
which indicates that the acetonitrile electronic interaction with the silicon substrate might 
be stronger than that for 1-propanol.

\subsection{Density of States}  

\begin{figure} 
\begin{center}
\includegraphics[width=5.0in]{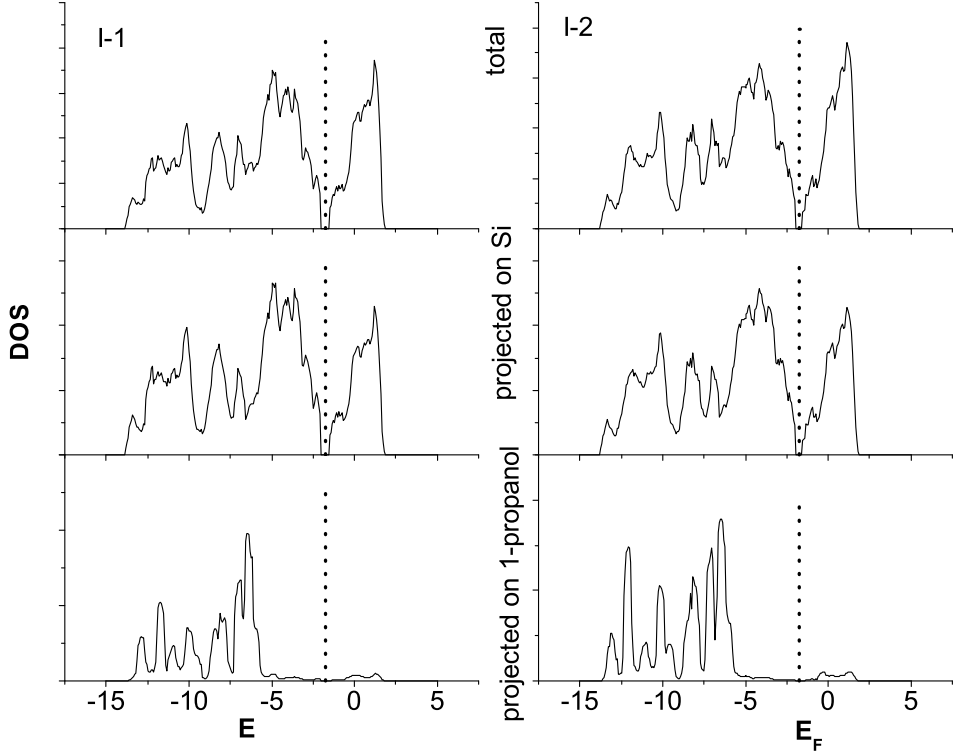}
\vspace{-1.0cm}
\end{center}
\caption{The electronic DOS and partial DOS projected on Si and 1-propanol
molecule for the physisorbed configurations I-1 and I-2 at 0.125 ML 1-propanol coverage.
The vertical dotted lines represent the positions of the Fermi level.}  
\label{8di-dos-physi} 
\end{figure}

\begin{figure} 
\begin{center}
\includegraphics[width=5.0in]{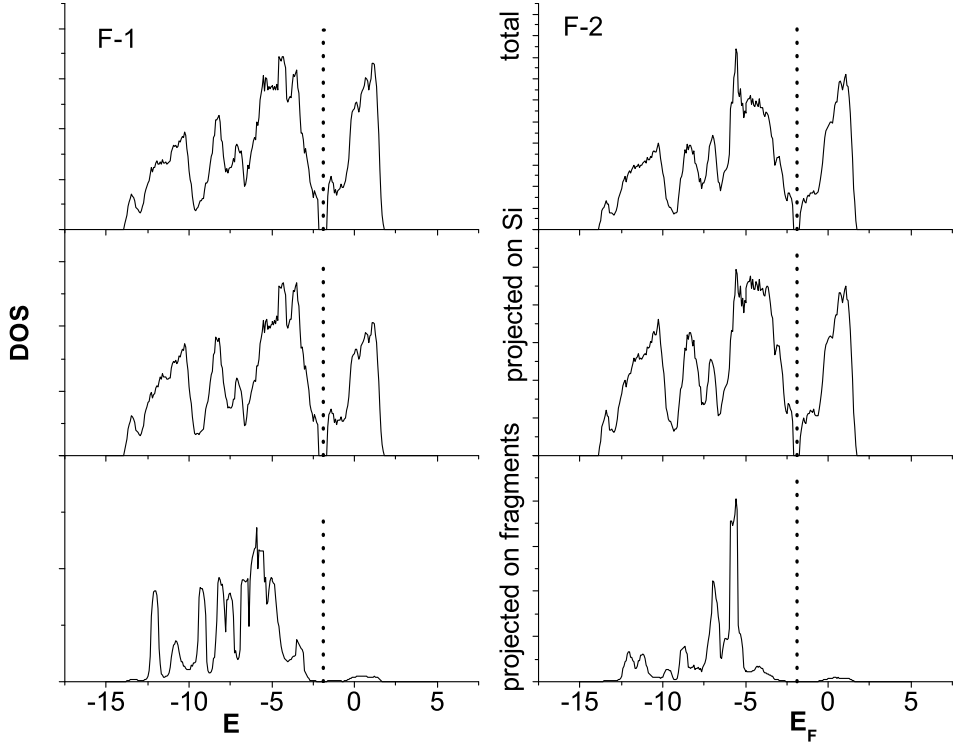}
\vspace{-1.0cm}
\end{center}
\caption{The electronic DOS and partial DOS projected on Si and 1-propanol
fragments for the chemisorbed configurations
F-1 and F-2 at 0.125 ML 1-propanol coverage. The vertical dotted lines represent the positions of the Fermi level.}  
\label{8di-dos-chemi} 
\end{figure}

The electronic density of states (DOS) as well as the partial DOS projected on Si atoms and the 1-propanol
molecule (or its fragments) for the physisorbed configurations I-1 and I-2, and 
the chemisorbed configurations
F-1 and F-2 are shown in Fig. \ref{8di-dos-physi} and Fig.\ref{8di-dos-chemi}.
The top, middle and bottom layers represent the DOS of the total slab,
the partial DOS projected on Si and the 1-propanol
molecule (or its fragments for chemisorbed structures), respectively.
The partial DOS projected on the 1-propanol molecule (or its fragments) is much lower in magnitude
than the DOS of the total slab for 0.125ML coverage. To show its features manifestly, 
we have rescaled this distribution (vertical axis) (see Figs. \ref{8di-dos-physi} and \ref{8di-dos-chemi}),
and the rescale factor is 10.

We consider the DOS with special emphasis on the peaks 
around the Fermi level and some characteristic peaks. 
The partial DOS projected on the Si atoms has almost the same profile 
as that of the DOS of the total slab, which
indicates that, for 0.125 ML coverage, the total DOS is dominated by the Si(001) surface states. 

In the following discussion, we focus on the two projected partial DOS distributions.
The main components of the peaks of the partial DOS distributions projected on the Si atoms and 1-propanol 
(or its fragments) are indicated in Table \ref{dospeak} for the configurations I-1, I-2, F-1 and F-2 

\begin{table}
\caption{The main components of the peaks of the partial DOS distributions projected on the Si atoms and 1-propanol
(or its fragments) for the configurations I-1, I-2, F-1 and F-2. The $3p$ represents
$3p_x$, $3p_y$ and $3p_z$.}
\begin{center}
\begin{tabular}{cccccccc}
\hline
\multicolumn{2}{c}{Configurations} & peak 1 & components & peak 2 & components &peak 3 & components\\
\multicolumn{2}{c}{~}& position & of peak 1 & position & of peak 2 & position & of peak 3\\
\hline
~~~~& partial DOS & -3.0 $\sim$-5.0ev & $3p$ & 1.4ev & $3p_x$,$3p_y$ &-11.2 $\sim$ -12.5ev & $3p$\\
~~~~& [Si] & ~~~~ & [Si] & ~~~~ & [Si] &~~~~& [Si]\\
I-1~& partial DOS &-11.7ev~&$2p_x$,$2p_y$ &-7.1ev~&$2p_z$ &-6.4ev&$2p_x$ [C3],\\
~~~~& [1-propanol] &~~~~& [O] &~~~~& [C2,C3] & ~~~~ & $2p_y$ [C1] \\
\hline
~~~~& partial DOS & -3.0 $\sim$-5.0ev & $3p$ & 1.4ev & $3p_x$,$3p_y$ & -11.2 $\sim$ -12.5ev & $3p$\\
~~~~& [Si] & ~~~~ & [Si] & ~~~~ & [Si] &~~~~& [Si]\\
I-2~& partial DOS &-12.1ev~&$2p_y$ &-7.1ev~&$2p_z$ &-6.5ev & $2p_y$ \\
~~~~& [1-propanol]&~~~~& [O]~&~~~~& [C2,C3] & ~~~~ & [C2,C3]\\
\hline
~~~~& partial DOS & -3.0 $\sim$-5.0ev & $3p$ & 1.4ev & $3p_x$,$3p_y$ & -11.2 $\sim$ -12.5ev & $3p$\\
~~~~& [Si] & ~~~~ & [Si] & ~~~~ & [Si] &~~~~&[Si]\\
F-1~& partial DOS &-9.3ev & $2p_z$ [C1], &-6.7ev & $2p_x$ [C3], & -6.1ev & $2p_y$ [C3],\\
~~~~& [H+alkoxy] &~~~~& $2p_x$ [O] &~~~~& $2p_y$ [O] & ~~~~ & $2p_z$ [C2]\\
\hline
~~~~& partial DOS & -3.0 $\sim$-5.0ev & $3p$ & 1.4ev & $3p_x$,$3p_y$ & -11.2 $\sim$ -12.5ev & $3p$\\
~~~~& [Si] & ~~~~ & [Si] & ~~~~ & [Si] &~~~~& [Si]\\
F-2~& partial DOS &-12.5ev& $2p_x$,$2p_y$ &-6.8ev & $2p_y$ [C1], &-5.5ev & $2p_x$,$2p_y$\\
~~~~& [OH+alkyl] &~~~~&[O]&~~~~& $2p_z$ [C3] &~~~~& [C2,C3]\\
\hline
\end{tabular}
\end{center}
\label{dospeak}
\end{table}

Comparing the partial DOS distributions projected on the 1-propanol molecule for the
configurations I-1 and I-2 (see Table \ref{dospeak}), we see that the second peak is the same for
both configurations,
but the first and third peaks are different which reflects the fact that the O-H bond direction for I-2 is by $180^{\circ}$ rotated with respect to
that of the configuration I-1. Comparing the DOS of the total slab,
the partial DOS projected on the Si atoms as well as on the 1-propanol molecule (see Figs. \ref{8di-dos-physi}
and \ref{8di-dos-chemi})
we find that near the Fermi level, the DOS is dominated by states that stem from
the Si(001) substrate, but at low energy (far below the Fermi level), 
the total DOS is modulated by the profile of the 1-propanol admixture (or its fragments). 

\section{Dependence on the level of coverage}  

Taking advantage of the slab approach, we will discuss in the following the dependence of the binding energy 
of the four basic configurations (I-1, I-2, F-1, F-2) on the coverage of the 1-propanol molecules. 
First, we will consider the basic configurations I-1, I-2, F-1 
and F-2 with the coverage levels 1.00ML, 0.5ML, 0.25ML and 0.125ML, which corresponds to one 
1-propanol molecule attached to one, two, four and eight dimers, respectively. Table \ref{bindingenergy} 
shows the binding energies of the four configurations of 1-propanol on the 
Si(001)-($2\times 1$) surface.

\begin{table}
\caption{Binding energies of the adsorbate on Si(001) in eV/per 1-propanol molecule
at four coverage levels}
\begin{center}
\begin{tabular}{ccccc}
\hline
Coverage~~~~~ &I-1~~~~~ & I-2~~~~~& F-1~~~~~& F-2~~~~~\\
\hline
0.125~~~~~ & 0.76~~~~~ & 0.72~~~~~   & 2.59~~~~~& 2.89~~~~~\\
0.250~~~~~ & 0.73~~~~~ & 0.70~~~~~    & 2.57~~~~~& 2.90~~~~~\\
0.500~~~~~ & 0.68~~~~~ & 0.67~~~~~    & 2.55~~~~~ & 2.91~~~~~\\
1.000~~~~~ & 0.41~~~~~ & 0.37~~~~~   & 2.34~~~~~ & 2.85~~~~~\\
\hline
\end{tabular}
\end{center}
\label{bindingenergy}
\end{table}

Table {\ref{bindingenergy}} shows that the binding energies per 1-propanol molecule for the physisorbed 
configurations I-1 and I-2 decrease with increasing coverage. This trend appears quite natural since increasing concentration 
of the adsorbed molecules on the Si(001) surface results in enhanced interaction between the molecules and hence 
weakens their bond with the substrate. The binding energy for the chemisorbed structure F-1 decreases 
with increasing coverage too. This may be related to the fact that the alkoxy fragment has similar transverse 
dimensions as the 1-propanol molecule. However, the binding energy for the chemisorbed configuration F-2 
exhibits very little change with the variation of the coverage.  This observation is ascribed to the 
strengthened interaction between the alkyl fragments and OH groups on different dimers with increasing 
coverage, counteracting the destabilization trend due to the enhanced alkyl 
density.

The dependence of the energy barriers on the 1-propanol coverage in the interval  [0.25 ML, 1.0 ML] is 
illustrated by Table \ref{barriercov}, which contains the energy barriers $E_b$ and transition state energy 
levels $E_{TS}$ with respect to the energy of 
1-propanol and Si(001) in isolation from each other. Two processes correspond to O-H bond and C-O bond scission. 

\begin{table}
\caption{Energy barriers $E_b$ and transition state energy levels $E_{TS}$, for the dissociation processes
I-1$\rightarrow$ F-1 and I-2$\rightarrow$ F-2 at four levels of surface coverage. The reference for the indicated energy values is the energy of the separated subsystems. }
\begin{center}
\begin{tabular}{cccccc}
\hline
Coverage~&I-1$\rightarrow$F-1($E_b$)~~&I-1$\rightarrow$F-1($E_{TS}$)~~
&I-2$\rightarrow$F-2($E_b$)~~&I-2$\rightarrow$F-2($E_{TS}$)\\
\hline
0.125&0.05&-0.70&1.34&0.62\\
0.250&0.02&-0.72&1.34&0.63\\
0.500&0.02&-0.65&1.32&0.65\\
1.000&0.05&-0.36&1.21&0.84\\
\hline
\end{tabular}
\end{center}
\label{barriercov}
\end{table}

Here we point out that  the 1-propanole molecules are placed on the surface uniformly, and all the molecules 
dissociate simultaneously.
From $E_b$ values for the O-H and C-O bond rupture in Table \ref{barriercov}, 
we find that the energy barriers 
for the O-H bond scission are only slightly affected by the level of coverage. However, the energy 
barriers for the C-O bond breaking ($E_b$)  
decrease with the increasing coverage.  

From Table \ref{bindingenergy}, the bonding of the chemisorbed structure F-1 (O-H bond scission ) 
is weakened as the coverage increases, and Table \ref{barriercov} reveals that the energy barrier with respect to the 
O-H bond rupture ($E_b$)
becomes higher at 1.00ML coverage. Thus, the probability of O-H bond scission is somewhat reduced at this level.  
On the other hand, the binding energy for the C-O bond cleavage changes very little as the coverage is varied, 
while the energy barrier with respect to the rupture of the C-O bond ($E_b$) has its minimal value at 1.00ML. 
This suggests that at a high coverage level, a small amount of C-O cleavage might occur, as supported by 
the experimental observation at high surface coverage \cite{czc}.

\begin{figure} 
\begin{center}
\includegraphics[width=5.0in]{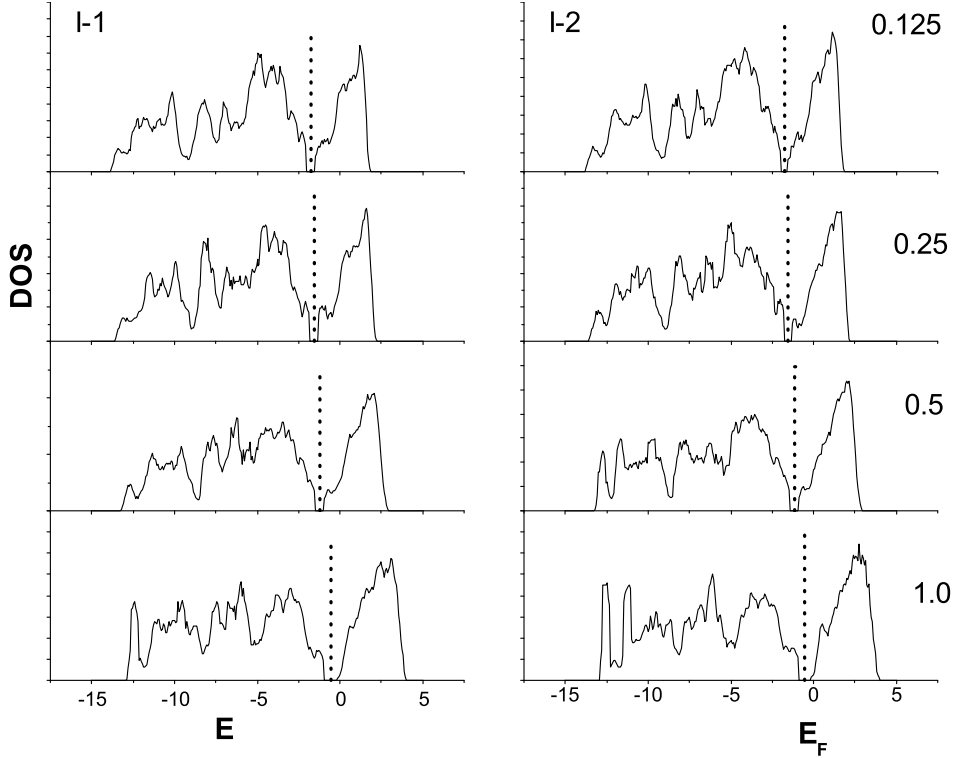}
\vspace{-1.0cm}
\end{center}
\caption{The DOS of the full slab for the physisorbed configurations I-1 and I-2 at 0.125,0.25,
0.5 and 1.0 ML.}  
\label{physi-dos-cov}
\end{figure}

\begin{figure} 
\begin{center}
\includegraphics[width=5.0in]{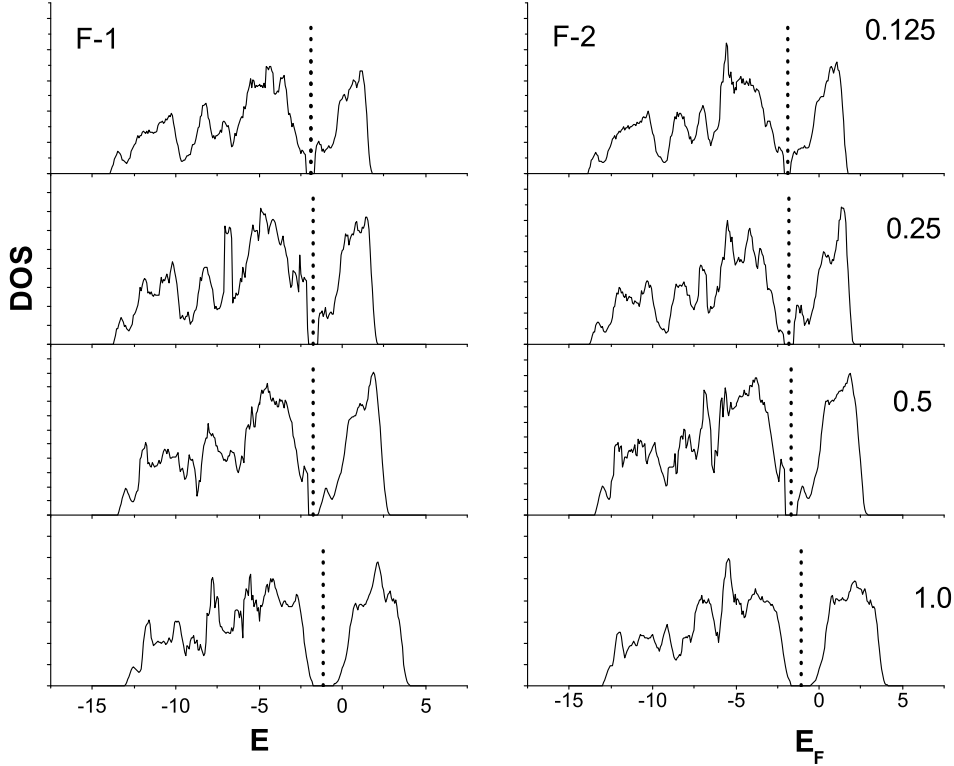}
\vspace{-1.0cm}
\end{center}
\caption{The DOS of the full slab for the chemisorbed configurations F-1 and F-2 at 0.125,0.25,
0.5 and 1.0 ML.}  
\label{chemi-dos-cov}
\end{figure}

The DOS distributions of the 1-propanol molecules 
(or its fragments) at the coverage levels considered in this work are shown in Fig. \ref{physi-dos-cov}
for the physisorbed configurations I-1 and I-2, and  Fig. \ref{chemi-dos-cov}
for the chemisorbed configurations F-1 and F-2.

The partial DOS projected on the 1-propanol molecule for the physisorbed I-1
structure at 0.125ML exhibits peaks at -6.4 ev and -7.1 ev 
which are traced back to 2$p$ orbitals of the carbon atoms C1,C2 and C3, and the peak
at -11.7 ev originates from the 2$p_x$ and 2$p_y$ orbitals of the oxygen atom.
 With increasing coverage
the peaks at -6.4 eV, -7.1 eV and -11.7 eV 
are found to grow if the substrate peak between -3.0 ev and -5.0 eV is taken as reference, which is the expected behavior upon 1-propanol deposition enhancement. This
conclusion applies for all the physisorbed and chemisorbed configurations.

For more quantitative analysis of the DOS distributions,
we examined the dependence of the energy gap on the 1-propanol coverage 
(or its fragments). Table \ref{gap} shows the obtained values
for the physisorbed configurations I-1 and I-2, and the chemisorbed configurations 
F-1 and F-2 with 0.125ML, 0.25ML, 0.5ML and 1.0ML.

\begin{table}
\caption{Energy gaps $\Delta E$ (eV) of the Si(001)-($2\times 1$) pure surface
compared to those of the 1-propanol adsorption structures
I-1,I-2,F-1,and F-2 on Si(001) at four levels of coverage}
\begin{center}
\begin{tabular}{cccccc}
\hline
Coverage&Si surface &I-1&I-2&F-1&F-2\\
\hline
0.125&0.46&0.53&0.40&0.53&0.40\\
0.250&0.46&0.53&0.47&0.60&0.60\\
0.500&0.46&0.53&0.53&0.80&0.73\\
1.000&0.46&0.80&0.73&1.27&1.20\\
\hline
\end{tabular}
\end{center}
\label{gap}
\end{table}

The energy gap for the Si(001)-($2\times 1$) surface is 0.46 ev, see Table \ref{gap}, which
is in keeping with experiment (the 
corresponding experimental value is about 0.6 ev\cite{wm}). The local DFT and GGA procedures tend to
underestimate the energy gap of semiconductors by up to 25$\%$ \cite{osg}.
Table. \ref{gap} shows that the energy gaps increase with the level of coverage, which
reflects a growing degree of saturation of the silicon dangling bond as induced by the 1-propanol molecules. As a consequence of a higher number of oxygen atoms attached to the surface, and, by the same token,  of dative bonds (for
the physisorbed configurations I-1 and I-2) or covalent bonds (for
the chemisorbed configurations F-1 and F-2) between oxygen atoms and silicon atoms,
the surface turns increasingly insulating, i.e., the energy gap widens.

\section{Room temperature Molecular Dynamics calculations}  

\begin{figure} 
\begin{center}
\includegraphics[width=3.0in]{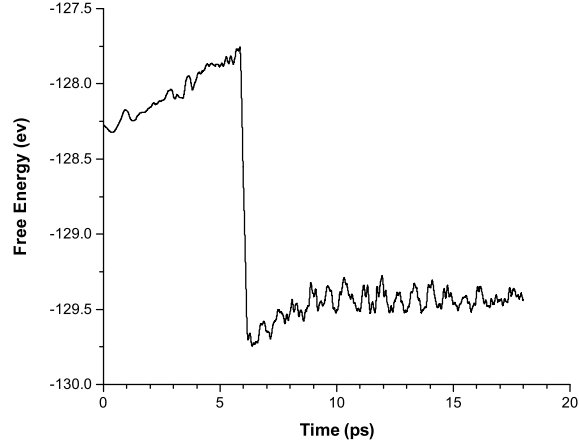}
\vspace{-0.9cm}
\end{center}
\caption{Time variation of the free energy in the MD simulation.
An average over every 300MD steps has been taken to filter out
high thermal frequency components.}
\label{free}
\end{figure}

\begin{figure} 
\begin{center}
\includegraphics[width=3.0in]{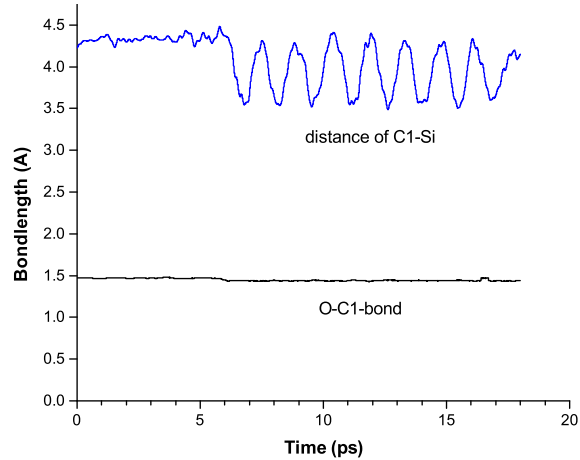}
\vspace{-0.9cm}
\end{center}
\caption{(Color online) Time variations of the Si-C1 distance and O-C1 bondlength in the MD evolution.
An average over every 300MD steps has been taken to filter out
high thermal frequency components.}
\label{cbond}
\end{figure}

\begin{figure} 
\begin{center}
\includegraphics[width=3.0in]{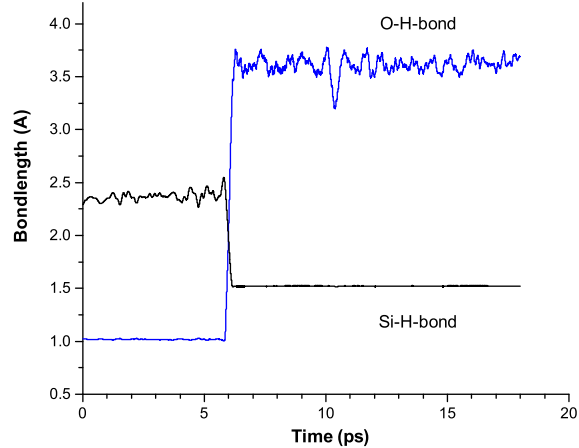}
\vspace{-0.9cm}
\end{center}
\caption{(Color online) Time variation of the O-H and Si-H bondlength in the evolution that O-H bond is breaking and
Si-H bond is forming.An average over every 300MD steps has been taken to filter out
high thermal frequency components.}
\label{hbond}
\end{figure}

The energy barrier computations have shown that the physisorbed 1-propanol molecule reacts with the 
Si(001)-($2\times 1$)
surface by cleavage of the O-H bond. Since the zero temperature transition state analysis may not be able to
access all of the relevant phase space volume, we perform a finite temperature ab initio MD simulation 
to take into account additional
possible reactions at T = 300K. The $1\times 2$ cell is adopted to carry out the MD simulation 
(the $2\times 2$ cell 
was used as well, and the results from both approaches were found to agree).
 In the finite temperature MD calculations
all atoms, including the passivating H atoms at the bottom of the slab, are allowed to
move. In this manner, a large temperature gradient can be avoided. Lattice parameters are
expanded according to the temperature under study using the experimental 
thermal expansion coefficient in order to prevent the lattice
from experiencing internal thermal strain\cite{scb}.
The starting configuration is the physisorbed one I-1,
see Fig. \ref{physisorption}, the O-H and O-C bond lengths are
1.01 $\AA$ and 1.48 $\AA$, respectively. The I-1 physisorbed
structure is heated to 300K (room temperature) in 9000MD step (9.0 ps, i.e., each
step takes 1 fs), followed by
another 3000MD steps at 300K  to evolve the system under conditions of thermal equlibrium. 
 Displaying the free energy of the system
as a function of the evolution time, we assess if the system has reached its equilibrium. 
As illustrated by Figure \ref{free}, the free
energy fluctuates very little after 10ps, which shows the system is at 
equilibrium. In Figs. \ref{free}, \ref{cbond} and \ref{hbond}, every data point represents
an average result over an interval of 300MD steps. In this way, high 
frequency components due to thermal motion\cite{lwh} are filtered out.

We consider the time variation of the O-C1 and Si-C1 bond lengths in the MD calculation, 
which are represented in Fig. \ref{cbond}. It is seen that 
the O-C1 bond is not ruptured in the process of the simulation. 
For times shorter than 6ps, the distance between the C1 and the up Si atom fluctuates around
4.25 $\AA$; between T = 6ps and T = 7ps, it reduces by 1 $\AA$, and after
T = 7ps, it oscillates around 4.0 $\AA$, which shows that no bond between C1 and
the up Si is formed. 
This behavior rules out the chemisorbed configuration F-2 as an equilibrium
structure.

On the other hand, the time variation of the O-H and  Si-H bond lengths in the MD simulation, as drawn in 
Fig. \ref{hbond}, illustrates that before T = 6ps, the O-H bond length is about 1.01 $\AA$ 
and the Si-H bond length oscillates around 2.25 $\AA$. In the period of 
6ps $\sim$ 7ps, the O-H bond length elongates up to 3.75 $\AA$,
and the Si-H bond length shortens to 1.48 $\AA$. 
This marked change indicates a transition from the metastable physisorbed
phase I-1 to the ``stable" chemisorbed phase F-1.  The characterization as stable for
the F-1 structure only makes sense 
at room temperature, since the chemisorbed
phase F-2 is much more stable than F-1 at still higher temperature.  After 7ps,
the O-H bond length oscillates with decreasing amplitude, and 
the Si-H bond length reaches its equilibrium value of 1.48 $\AA$. 
Fig. \ref{hbond} shows that O-H bond scission occurs and
the Si-H bond forms between T = 6ps and
T = 7ps. The equilibrium structure 
is the chemisorbed configuration
F-1, i.e., the O-H bond is broken (see Fig.\ref{chemisorption}), which is consistent with the energy barrier
calculation at zero temperature.

Inspection of the MD simulation results shows that the O-H bond is
broken. The H atom is detached and reattaches to the ``up"
silicon atom of the same dimer (dimer A2) to form a new H-Si bond. After 7ps, all
atoms oscillate around their stable equilibrium positions. Another method of performing the MD simulations
consists in setting an initial temperature T equal to 300K without any heating and letting the system evolve at this temperature. Following this avenue, we arrive at the same conclusions as reported above.

\section{Summary} 
 
We have performed a study on the physisorption and chemisorption of 1-propanol molecules on the Si(001)-($2\times 1$) 
surface from first principles.  Phenomena related to the geometric, electronic, energetic and fragmentation pathways have been investigated within three-dimensional periodic boundary conditions. Specifically, we have shown that the 1-propanol molecule initially 
interacts with the Si surface through formation of a dative bond. Subsequently, the physisorbed 1-propanol molecule reacts 
with the surface by cleavage of either the O-C or the O-H bond. The O-C bond cleavage is thermodynamically stable, 
but the O-H bond cleavage is kinetically favored. We characterized the geometric modification of the 
Si(001)-($2\times 1$)
surface in response to physisorption as well as chemisorption, 
which cannot be described by use of a single-dimer cluster model. 

We have first calculated the band structure and the DOS for four 
configurations, demonstrating that the occupied bands within the fundamental band gap of the silicon are composed of the up Si atoms,
and the unoccupied bands originate from the down Si atoms. No conduction band within the fundamental band gap could be 
associated with Si-O bonding. This feature distinguishes the present case from that of acetonitrile adsorption 
on the silicon surface\cite{mop}. For acetonitrile adsorption, 
a conduction band within this gap has been assigned to Si-N bonding, 
which shows that the acetonitrile
electronic interaction with the silicon substrate might be stronger than that of 1-propanol. The peaks 
around the Fermi level and other peaks related to adsorption for the DOS and partial DOS distributions were discussed. 
It has been shown that 
the DOS near the Fermi level is dominated by the states from the
Si(001) surface, but at low energy (far below the Fermi level) the DOS is modulated by the 1-propanol contribution, or that of its fragments.

We have analyzed 
the dependence of the various properties (binding energy,
energy barrier, density of states (DOS), energy gap) for the configurations I-1, I-2, 
F-1 and F-2 on the coverage 
levels 0.125ML, 0.25ML, 0.5ML and 1.0ML. From this research, the binding energies of the physisorbed 
configuration I-1, I-2 and chemisorbed configuration F-1 decrease with increasing coverage. This trend appears plausible 
since increasing concentration of 1-propanol molecules on the Si(001) surface results in enhanced repulsion between 
the molecules and hence destabilization. However, the binding energy for the chemisorbed configuration F-2 is found 
to be rather insensitive to the variation of the coverage level.
The energy barrier 
with respect to O-H bond scission at four levels of coverage changes slightly, and reaches its maximum within 
the [0.25 ML, 1.0 ML] interval at 1.00ML. However, the energy barriers with respect to the C-O bond 
rupture ($E_b$) decrease with
increasing coverage. Thus, we found the bonding of the chemisorbed structure F-1 (O-H bond scission ) 
destabilized at higher coverage, while the energy barrier is highest at 1.00 ML, suggesting a reduced probability of O-H bond 
scission at the 1.00ML level.  On the other hand, the binding energy for the C-O bond breaking 
structure changes very little with the level of coverage, while the energy barrier with respect to the C-O 
bond rupture ($E_b$)  
adopts its minimal value at 1.00 ML. One concludes that conditions of high coverage favor C-O cleavage, 
and a small admixture of this chemisorption channel may be observable at 1.0 ML \cite{czc}.
The DOS showed that upon increasing 1-propanol deposition, the peaks due to the 1-propanol molecular orbitals
grow in intensity, and the substrate features diminish. Also, with enhanced coverage, the energy gaps widen which
indicates increasing saturation of the silicon dangling-bonds by the 1-propanol molecules. 
The planar 1-propanol density on the Si(001) surface thus represents a 
parameter that allows to alter the nature of the surface from semiconducting to insulating.

Finally, recording the time variation of the O-H and Si-H bond lengths by means of ab initio MD simulation 
demonstrated that the O-H bond length is spontaneously ruptured at room temperature, and 
the dissociated H atom forms a Si-H bond. The final equilibrium structure 
at room temperature is the chemisorbed configuration F-1. The observed O-H bond rupture 
is in accordance with the energy barrier calculation at zero temperature.

 \section*{Acknowledgments}  We thank A. Arnaldsson, A. Ciani, G. Henkelman, R. Miott and P. Silvestrelli 
for the correspondence. 
This work is supported by the National Science Foundation through the grants HRD-9805465, 
NSFESP-0132618 and DMR-0304036, by the National Institute of Health through the grant S06-GM008047, 
and by the Army High Performance Computing Research Center under the auspices of Department of the Army, 
Army Research Laboratory under Cooperative Agreement No. DAAD 19-01-2-0014. 
   
\end{document}